\documentclass[prd,reprint,superscriptaddress]{revtex4-2}

\usepackage{amssymb}
\usepackage{graphicx}
\usepackage{slashed}
\usepackage{color}
\usepackage{amsmath}
\usepackage{xspace}
\usepackage{xcolor}
\usepackage{soul}
\usepackage{appendix}
\usepackage{multirow}

\newcommand{\nopieft}{\mbox{$\slashed{\pi}$}EFT }

\newcommand{\be}{\begin{equation}}
\newcommand{\ee}{\end{equation}}

\newcommand{\bra}{\langle}
\newcommand{\ket}{\rangle}
\newcommand{\rvec}{\ensuremath{\boldsymbol{r}}}
\newcommand{\avgL} {\bra L \ket}
\newcommand{\fm}{\,\mathrm{fm}}

\begin{document}

\title{
Perturbative application of next-to-leading order pionless EFT for $A\le3$ nuclei in a finite volume
}

\author{Tafat Weiss-Attia}
\affiliation{The Racah Institute of Physics, The Hebrew University, 
               Jerusalem 9190401, Israel}

\author{Martin Sch\"{a}fer}
\email{m.schafer@ujf.cas.cz}
\affiliation{Nuclear Physics Institute of the Czech Academy of Sciences, 
             Rez 25068, Czech Republic}

\author{Betzalel Bazak}
\email{betzalel.bazak@mail.huji.ac.il}
\affiliation{The Racah Institute of Physics, The Hebrew University, 
               Jerusalem 9190401, Israel}

\date{\today}

\begin{abstract}
Lattice quantum chromodynamics (LQCD) calculations with physical pion mass would revolutionize nuclear physics by enabling predictions based on the fundamental theory of the strong force.
To bridge the gap between finite-volume LQCD results and free-space physical observables, two primary extrapolation methods have been employed so far. The traditional approach relies on the L\"{u}scher formula and its extensions, while a recent alternative employs effective field theories (EFTs) fitted directly to the finite volume data. 
In this study, we fit pionless EFT with perturbative inclusion of the next-to-leading order to finite-volume energies generated from a phenomenological $NN$ interaction. The theory is then used to extrapolate the finite-volume results into free space as well as to predict new few-body observables.
As a benchmark, we also apply the L\"{u}scher formalism directly to the finite-volume data.
Through a comprehensive analysis, we explore the characteristics of order-by-order predictions of the pionless EFT fitted within a finite volume,
investigate the limitations of the different extrapolation techniques used, and derive recommended box sizes required for reliable predictions.
\end{abstract}


\maketitle

\section{Introduction}

In the realm of nuclear physics, future \emph{ab-initio} predictions for low energy observables would ideally be based on lattice chromodynamics (LQCD), a lattice gauge theory of quarks and gluons at low energies \cite{Wilson, LQCD23, Mainz21, PACS, CalLat}. 
LQCD involves extensive numerical calculations performed on a finite space-time lattice, and in the limit of an infinitely large lattice size and infinitesimal spacing, it provides an exact solution to QCD. 
However, practical LQCD calculations are often performed with a small lattice size, where finite-volume effects come into play, and therefore require extrapolation of the results into free-space observables to give them physical meaning. At the moment, calculations with physical quark masses are not available; however, there is ongoing progress in approaching the physical point, see for example Refs.~\cite{HALQCD1, HALQCD2}.

Several decades ago, L\"{u}scher developed a method for extracting free space observables from finite-volume spectra of two-body systems confined in a box of size \mbox{$L\times L \times L$} \cite{Luscher_bound, Luscher_scatt}.
This approach is based on the assumption of scale separation, where the potential range $R$, the possible binding momentum $\kappa$, and the box size $L$ satisfy the conditions $R \ll \kappa^{-1} \ll L$. Nonetheless, this assumption may not always hold for all systems, considering the characteristic scales of the nuclear interaction and the computational limitations of current LQCD calculations. For instance, the long-range part of the nuclear interaction has a range of approximately $R \approx 2\ \fm$, while the typical deuteron binding momentum is around $\kappa^{-1} \approx 5\ \fm$. Moreover, due to computational constraints, the box size in LQCD calculations is typically fixed at about $L \approx 5\ \text{fm}$. As the box size decreases, it becomes essential to include exponentially suppressed corrections \cite{DavSav11, SatBed07}. Ref. \cite{DavSav11} demonstrates that incorporating corrections to L\"{u}scher's asymptotic formula allows extraction of the deuteron binding energy with high precision when using box sizes of $ L\gtrsim 10 \fm$. 

Finite-volume corrections in three-body systems were studied in Refs. \cite{BriDav13, HanSha14, HanSha15, MeiRioRus15, BriHanSha17, HamPanRus17, HamPanRus17B, MaiDoring17}. So far, theoretical studies beyond three particles involve only systems where dominant finite-volume corrections might be described through the separation of $N$-body systems into two subclusters \cite{KonLee18,Kon20,Yu23}.

An alternative approach was proposed in Ref. \cite{ElBazBar}, which suggests the use of effective field theory (EFT; see, e.g., \cite{HamKonKol20} for a recent review). Here, the few-body Schrödinger equation is solved in the same finite volumes used in the LQCD calculations, and an EFT is fitted to reproduce the corresponding LQCD results. Subsequently, the infinite volume quantities are obtained by solving this EFT in free space. Specifically, a pionless effective field theory ($\slashed{\pi}$EFT) \cite{KapSavWis, BedHamKol, Kolck, BirMcRich}, where the nucleons are the only degrees of freedom, was utilized in Ref. \cite{ElBazBar} to analyze the nuclear spectrum obtained by the NPLQCD collaboration at a pion mass of $m_\pi \approx 806$ MeV \cite{NPLQCD1}. In later works, \nopieft was employed to calculate relevant matrix elements and analyze other LQCD results \cite{DetShan, NPLQCD2, DavKad22, DetRomShan}.

One of the most important features of EFT is its ability to systematically improve results by incorporating order-by-order corrections. The majority of studies use only leading order (LO) \nopieft to extrapolate LQCD results \cite{ElBazBar, DetShan, NPLQCD2, DavKad22, DetRomShan}. While the LO is to be resummed, the subleading range corrections of \nopieft are to be included perturbatively to ensure a properly renormalized theory. In free space, higher order terms were perturbatively included in nuclear three-body systems up to N$^2$LO \cite{Vanesse13} and in four- and five-body nuclear systems up to NLO \cite{SB23,BSBB23}. In finite-volume, the widely used particle-dimer formalism faces numerical challenges, once perturbative effective range corrections are included in the dimer propagator. As a result, an alternative scheme for effective range corrections was introduced \cite{EHR21,EHR23}. Recently, Ref. \cite{DetRomShan} applied NLO \nopieft to analyze the LQCD spectra of two-nucleon systems \cite{LQCD23}. In this work, the authors follow naive power-counting and implement at NLO all interaction terms with quadratic momenta non-perturbatively. Consequently, the theory faces renormalization issues due to the Wigner bound \cite{Wigner, Phillips97}, the renormalization group invariance can not be verified since the large cutoff limit can not be reached, and the model independence is obscured.

In this study, we employ the power-counting scheme as used in Refs.~\cite{Vanesse13,SB23,BSBB23}. The NLO contains effective range corrections, which are included through the perturbative insertion of two-body momentum-dependent $s$-wave terms. This allows us to access both the cutoff dependence of our two- and three-body \nopieft results as well as finite-volume effects induced by the fit of \nopieft low-energy constants (LECs) to finite-volume spectra. Calibration of LECs using finite-volume energies requires an accurate solution of the few-body Schrödinger equation in a box with periodic boundary conditions. Here we use the correlated Gaussian-based stochastic variational method (SVM), which has shown its ability to accurately capture the finite-volume effects \cite{YinBlum,YarBazSchBar}, and serves as a reliable tool for fitting \nopieft to LQCD data \cite{ElBazBar,DetShan,NPLQCD2,DavKad22,DetRomShan}. Specifically, we use the implementation from Ref. \cite{ElBazBar}, which has been further optimized and improved in Ref. \cite{YarBazSchBar} to obtain efficient and accurate computations in a periodic box, and which has been generalized in this work to include NLO terms as well. Due to the lack of nuclear LQCD results at physical pion mass, we artificially generate LQCD-like data. To this end, we assume that the nuclear interaction is fully described by the phenomenological Minnesota potential \cite{Minn}. The extracted free-space LO and NLO \nopieft results are then compared to the corresponding values calculated directly from the Minnesota potential. 
We observe a significant improvement in the results obtained using NLO  \mbox{\nopieft} in comparison to the LO results, indicating that the NLO corrections effectively account for finite volume effects. In most of the investigated systems, the NLO \mbox{\nopieft} yields predictions with high accuracy when utilizing box sizes $L \gtrsim 7\fm$.

The paper is organized as follows. Section II presents the L\"{u}scher extrapolation formulas, which are suitable for both bound states and scattering states.
Section III describes the \nopieft framework. Section IV presents the numerical tools used to solve the few-body problem, focusing on the implementation of periodic boundary conditions. An explanation of the fitting procedure for the EFT LECs is also included in this section. Our results are presented in Section V, followed by conclusions in Section VI. Finite-volume energies of two- and three-body systems calculated using Minnesota potential are listed in Appendix~A. In Appendix~B we include complementary \nopieft results.

\section{L\"{u}scher methods}

The common framework for processing the LQCD finite-volume data was introduced by L\"{u}scher \cite{Luscher_bound, Luscher_scatt}.
L\"{u}scher established the connection between the finite-volume spectra of a two-body system, confined in a box with periodic boundary conditions, and the physical observables of the same system in free space, specifically the binding energy and phase shifts.

The first-order finite-volume correction to the binding energy of an $s$-wave two-body bound state in a box of size $L$, $\Delta B_2 = B_2(L) - B_2^\mathrm{free}$,
is given by \cite{Luscher_bound}
\be\label{LuschBoundform}
 \Delta B_2 = \frac{6 \kappa_2 |\mathcal{A}_2|^2}{\mu _2 L} e^{-\kappa_2 L} +O(e^{-\sqrt{2}\kappa_2 L}),
\ee where $B_2^\mathrm{free}$ is the binding energy of the two-body system in free space and $B_2(L)$ is the corresponding binding energy in a finite volume with a box size $L$. Here we set $\hbar = 1$, $\mu_2$ is the reduced mass, $\kappa_2 = \sqrt{2\mu_2B_2^\mathrm{free}}$ is the binding momentum, and $\mathcal{A}_2$ is the dimensionless two-body asymptotic normalization coefficient (ANC).

A generalization of Eq.~(\ref{LuschBoundform}) for an $N$-body bound state was given in Ref. \cite{KonLee18}. Assuming that the lowest threshold of the system is a breakup into two subclusters and these subclusters can be treated as point-like particles, the leading finite-size correction to the free-space $N$-body binding energy $B_N^{\rm free}$ is
\be\label{NbodBound}
 \Delta B_N = C_N \frac{6 \kappa _N |\mathcal{A}_N|^2}{\mu _N L} e^{-\kappa_N L} +O(e^{-\sqrt{2}\kappa_N L}).
\ee Here, $\mu_N$ stands for the reduced mass of the subclusters,
$$\kappa _N = \sqrt{2\mu_N(B_N^{\mathrm{free}} - B_1^{\mathrm{free}} - B_2^{\mathrm{free}})}$$
is the binding momentum calculated from the free-space binding energies of the two subclusters
$B_1^{\mathrm{free}}$ and $B_2^{\mathrm{free}}$, $\mathcal{A}_N$ is the corresponding $N$-body ANC, and $C_N$ is a combinatorial factor counting the number of partitions of $N$ identical particles into these two subclusters.

A formula suitable for the analysis of two-body scattering states was given in Ref.~\cite{Luscher_scatt}. For a range of interaction $R$ smaller than the size of the box, $R<L$, the free-space scattering amplitude in a specific partial wave can be related to the energy spectra of two particles in a finite volume, with periodic boundary conditions and a certain cubic symmetry. Assuming that all partial waves higher than the $s$-wave can be neglected, the $s$-wave phase shifts $\delta_{0}(k)$ can be determined from the corresponding $A_1^+$ finite-volume energies by applying the equation \cite{Luscher_scatt}
\be \label{LuschScattForm}
  k \cot\delta_{0}(k) = \frac{1}{\pi L} S \left[\left(\frac{Lk}{2\pi}\right)^2 \right].
\ee Here $k$ is the relative momentum corresponding to the finite-volume energy $E=k^2/m$, with nucleon mass $m$. $S(\eta)$ is given by the regularized sum \cite{BeanBedParSav}
\be \label{S(eta)}
 S(\eta) \equiv \lim_{\Omega \to \infty}
 \left( \sum_{|\boldsymbol{j}|\in\mathbb{Z}^3}^{\Omega} \frac{1}{\vert\boldsymbol{j} \vert ^2 - \eta} - 4\pi \Omega \right),
\ee
running over all integer three-vectors $\boldsymbol{j}$ such that $|\boldsymbol{j}|<\Omega$. The next lowest partial wave that can also affect the $A_1^+$ finite-volume energy spectrum is the $g$-wave. At nuclear energies considered in this study, the $g$- or even higher partial-wave contributions to the spectrum are negligible, and their effect is not taken into account.

The scattering length and effective range can be obtained by fitting the extracted $s$-wave phase shifts with the effective range expansion (ERE)
\be \label{eq: ERE }
k \cot \delta_{0} (k) = -\frac{1}{a_{0}} + \frac{1}{2} r_{0} k^2 + O(k^4).
\ee
Using the scattering parameters, the two-body binding energy can be approximately calculated,
\begin{equation}\label{b_e_scatt_params}
 B_2 \approx \frac{1}{m a^2_{0}} \left( 1- \frac{r_{0}}{a_{0}}\right)^{-1}\,.
\end{equation}

\section{\nopieft up to NLO}

An alternative approach to utilize LQCD results is to fit the \nopieft directly to LQCD finite-volume spectra \cite{ElBazBar, DetShan, NPLQCD2, DavKad22, DetRomShan}. At LO, the \nopieft potential contains contact interaction in each two- and three-body $s$-wave channels. After regularization, the contact interaction is smeared by a local Gaussian regulator 
\be \begin{aligned}
  \delta_{\Lambda}(r) &=  \frac{\Lambda^3}{8\pi^{3/2}}\exp[-\Lambda^2r^2/4] \\
  &= \frac{\Lambda^3}{8\pi^{3/2}} \prod_{\alpha \in\{x, y, z\} } \exp[-\Lambda^2(r^{(\alpha)})^2/4],
\end{aligned} \ee
where $\Lambda$ is the momentum cutoff. The regularized LO potential takes the form
\be \begin{aligned}
  \hat{V}^{(0)} &= \sum_{i<j} \left( C_{0}^{(0)}(\Lambda)~\hat{\mathcal{P}}^{[0, 1]}_{ij} +C_{1}^{(0)}(\Lambda)~\hat{\mathcal{P}}^{[1, 0]}_{ij} \right)\delta_{\Lambda}(\rvec_{ij}) \\&+ \sum_{i<j<k}\sum_{cyc}D^{(0)}_0(\Lambda)~\hat{\mathcal{Q}}^{[1/2,1/2]}_{ijk}\delta_{\Lambda}(\rvec_{ij})\delta_{\Lambda}(\rvec_{jk}),
\end{aligned} \ee
where $\rvec_{ij} = \rvec_i - \rvec_j$ is the relative coordinate between particles $i$ and $j$ and $\hat{\mathcal{P}}^{[S, I]}_{ij}$, $\hat{\mathcal{Q}}^{[S,I]}_{ijk}$ are projection operators into the respective two- and three-body spin-isospin $[S,I]$ channels. Here $C_{0}^{(0)}(\Lambda)$, $C_{1}^{(0)}(\Lambda)$, and $D_{0}^{(0)}(\Lambda)$ are the LO LECs which have acquired cutoff dependence after regularization.

Moving into the next EFT order, the NLO potential contains derivatives of the contact potential, as well as counter-terms chosen to verify that the fitted LO observables remain unchanged. The NLO potential takes the form 
\be \begin{aligned}
  \hat{V}^{(1)} & = \sum_{i<j} \left( C_{0}^{(1)}(\Lambda)~\hat{\mathcal{P}}^{[0, 1]}_{ij} +C_{1}^{(1)}(\Lambda)~\hat{\mathcal{P}}^{[1, 0]}_{ij} \right)\delta_{\Lambda}(\rvec_{ij})  \\&+ \sum_{i<j<k}\sum_{cyc}D^{(1)}_0(\Lambda)~\hat{\mathcal{Q}}^{[1/2,1/2]}_{ijk}\delta_{\Lambda}(\rvec_{ij})\delta_{\Lambda}(\rvec_{jk}) \\&
  +\sum_{i<j} \left( C_{2}^{(1)}(\Lambda)~\hat{\mathcal{P}}^{[0, 1]}_{ij} +C_{3}^{(1)}(\Lambda)~\hat{\mathcal{P}}^{[1, 0]}_{ij} \right) \cdot \\& 
  \quad \cdot \left(\delta_{\Lambda}(\rvec_{ij})\overrightarrow{\nabla}^2_{ij}  + \overleftarrow{\nabla}^2_{ij} \delta_{\Lambda}(\rvec_{ij}) \right).
\end{aligned} \ee
At NLO, the adopted power-counting further includes a momentum-independent four-body force \cite{BazKirschetal,SB23}, which is not required in this $A \leq 3$ study.

The model independence of our approach is ensured by requiring that both the leading order and the next-to-leading order of \nopieft are properly renormalized to low-momentum data. While the LO potential is iterated, non-perturbative inclusion of the NLO momentum-dependent (derivative) terms leads to renormalization problems caused by the Wigner bound \cite{Wigner,Phillips97}. To maintain a properly renormalized EFT, the NLO potential is treated within first-order perturbation theory \cite{Kolck}.

The truncation of \nopieft at a certain order leads to a theoretical error that reflects higher order terms not accounted for in the EFT expansion. A rough LO and NLO error estimate is $(QR)$ and $(QR)^{2}$, respectively, where $Q$ is the typical momentum of an observable and $R \sim r_0$ is the interaction range.
Another possibility how to estimate the error is the residual cutoff variation which would be eventually removed by including all EFT orders \cite{Grießhammer}. 
In this work, we repeat \nopieft calculations for several values from a broad momentum range of $1.25~{\rm fm}^{-1}\leq \Lambda \leq 10~{\rm fm}^{-1}$ which allows to assess the cutoff convergence of our results as well as to estimate the truncation error.

In a periodic box of size $L$, the \nopieft potential must obey the periodic boundary conditions. Following Refs. \cite{YinBlum, ElBazBar}, this is achieved by summing over all possible translations of the box. As a result, the Gaussian regulator now has the form
\be \begin{aligned}
  \delta_{\Lambda, L}(r) = \frac{\Lambda^3}{8\pi^{3/2}} \prod_{\alpha \in \{ x, y, z\} } \sum_{q^ {(\alpha)}\in \mathbb{Z} } \exp[-\Lambda^2(r^{(\alpha)} - Lq^{(\alpha)})^2/4].
\end{aligned} \ee

\section{Methods}
\subsection{Stochastic Variational Method}
We solve the few-body Schrödinger equation by utilizing the SVM \cite{VarSuz98}. The total wave function $\Psi$ is expanded on a correlated Gaussian basis \cite{VarSuz95}
\be 
  \Psi = \sum_i c_i ~ \hat{\mathcal{A}} 
  \left \{ G_i(\rvec)\chi_{SM_S}\xi_{IM_I} \right\}.
\ee
$\hat{\mathcal{A}}$ denotes the anti-symmetrization operator over nucleons, and $\chi_{SM_S}$ and $\xi_{IM_I}$ are the spin and isospin parts of the wave function, respectively.
The spatial part is given by $G_i(\rvec)=\exp \left[-\frac{1}{2}(\rvec^T A_i \rvec) \right]$, where $\rvec^{T} = (\rvec_1, \rvec_2, \dots, \rvec_N)$ are the single-particle $N$-body coordinates, and $A_i$ is a symmetric positive-definite matrix of size $N\times N$, chosen stochastically. 
Finally, the variational coefficients $c_i$ and the associated bound-state energies are obtained by solving the generalized eigenvalue problem.

The total wave function in a box, $\Psi_L$, has to obey the periodic boundary conditions
\be
\Psi_L(\dots, \rvec_i, \dots) = \Psi_L (\dots, \rvec_i +\boldsymbol{m}_i L, \dots)
\ee
for arbitrary integer trio $\boldsymbol{m}_i = (m_i^{(x)}, m_i^{(y)}, m_i^{(z)})$.
The corresponding spatial part is then represented by a product of periodic correlated Gaussians in the $x,y,z$ directions
\be
G_L(\rvec) = \prod_{\alpha \in \{ x,y,z\} }G_{L_\alpha}(\rvec^{(\alpha)})
\ee
with 
$(\rvec^{(\alpha)})^T = (r_1^{(\alpha)}, r_2^{(\alpha)}, \dots, r_N^{(\alpha)})$.
Following Refs. \cite{YinBlum, YarBazSchBar}, the periodicity is achieved by summing over all box translations
\be
G_{L_\alpha}(\rvec^{(\alpha)}) = \sum_{\boldsymbol{n}^{(\alpha) }  }G_\alpha(A_\alpha; \rvec^{(\alpha)} - L\boldsymbol{n}^{(\alpha) } )
\ee
with $G_\alpha$ being a correlated Gaussian function
\be
G_\alpha(A_\alpha; \rvec^{(\alpha)} ) = \exp \left[-\frac{1}{2}( \rvec^{(\alpha)})^T A_\alpha \rvec^{(\alpha)} \right]
\ee
and 
$\boldsymbol{n}^{(\alpha) } = (n_1^{(\alpha)}, n_2^{(\alpha)}, ...,n_N^{(\alpha)})$, $n_i^\alpha \in \mathbb{Z} $.

It was shown in Ref. \cite{YinBlum} that it is beneficial to choose the $A_\alpha$ matrices that obey
\be
(\rvec^{(\alpha)})^T A_\alpha \rvec^{(\alpha)} = \sum_{i<j} \frac{(r^{(\alpha)}_i - r^{(\alpha)} _j )^2 }{b_{ij}^2},
\ee
where the stochastically selected parameters $\{b_{ij}\}_{i<j}$ reflect the relative distance between each pair of particles. Such basis functions are then invariant under center-of-mass shifts. As demonstrated in Ref. \cite{YarBazSchBar}, the center-of-mass coordinate can be integrated out in finite-volume calculations, further simplifying the numerical computations and eliminating center-of-mass excitations.

\subsection{Fitting the LECs \label{fitting}}

In place of genuine physical LQCD results, we use artificially generated finite-volume data. To this end, we apply the SVM to calculate the $A_1^+$ spectra of two and three nucleons in a periodic box, where the underlying nuclear interaction is described via the phenomenological $NN$ Minnesota potential \cite{Minn}. Using the exchange mixing parameter $u = 1$, this potential reproduces the experimental $np$ spin-triplet and $pp$ spin-singlet $s$-wave effective-range parameters. Furthermore, it was demonstrated that the Minnesota potential describes successfully the basic free-space properties of the deuteron, triton, and $\alpha$-particle \cite{Minn}. The generated finite-volume data are listed in Appendix~\ref{appA}.

\begin{figure}
  \includegraphics[width=8.6cm]{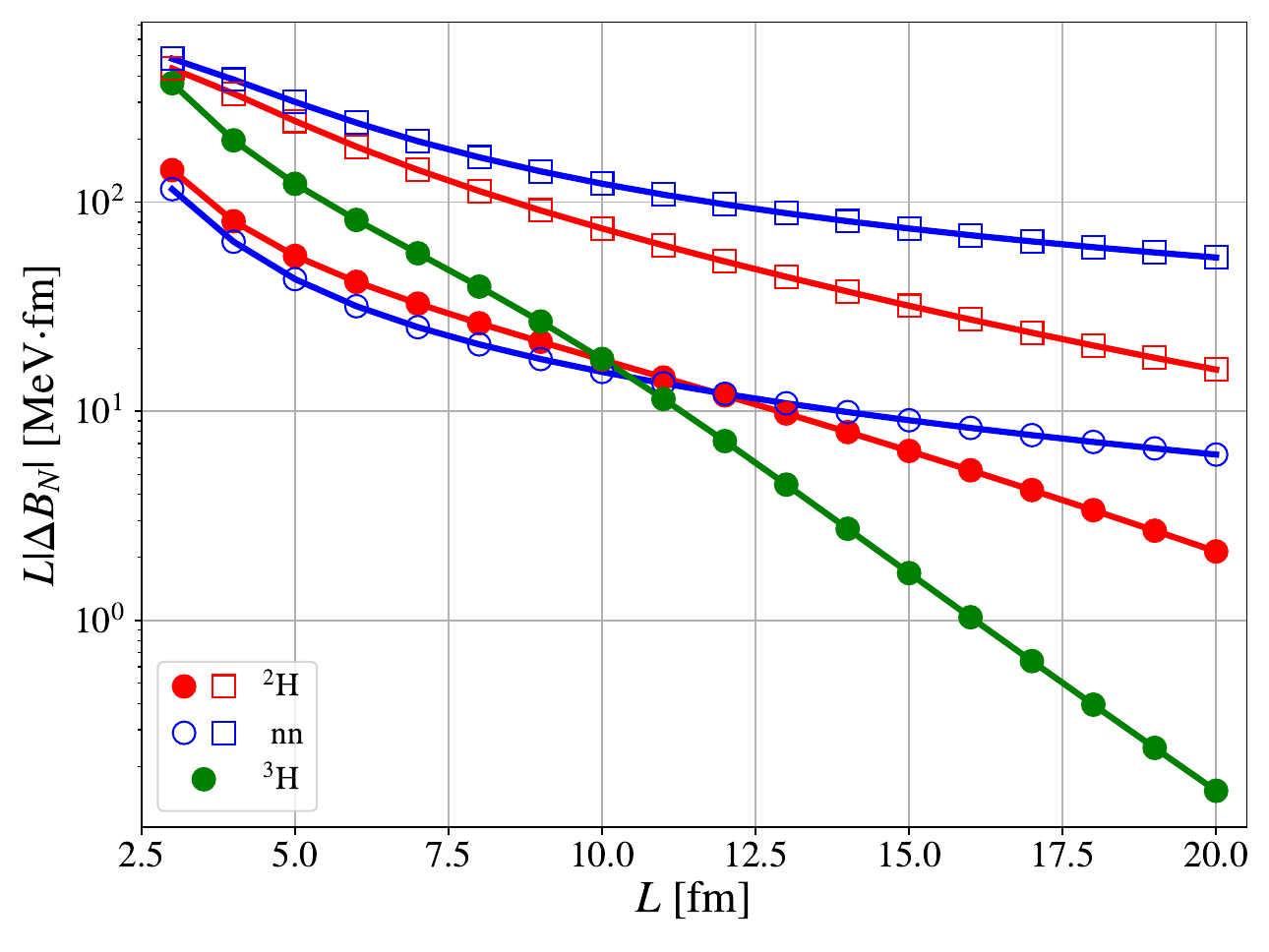}
  \caption{The finite-volume results used to mimic LQCD data in this work. The data were obtained by solving the Minnesota potential in a periodic box with different box sizes $L$. Shown are the absolute values of the finite volume energy shifts $|\Delta B_N|$, multiplied by the box size $L$, as a function of the box size for the deuteron (red),  dineutron (blue), and triton (green) channels. The circles correspond to the finite-volume ground states. For the two-body systems, the values of the first excited state are depicted as well (squares). The filled symbols indicate the finite-volume deuteron and triton ground states which in the the large $L$-limit approach the free-space bound states. 
  }
  \label{fig:Minn}
\end{figure}

Motivated by the L\"{u}scher bound-state formulas, Eqs. (\ref{LuschBoundform}) and (\ref{NbodBound}), we show in Fig.~\ref{fig:Minn} the calculated finite-volume corrections $L|\Delta B_N|$ to the free-space Minnesota deuteron and triton binding energies. The corrections are plotted as a function of $L$ on a logarithmic scale. The linear behavior then reveals the range of applicability of the L\"{u}scher formulas, $L \gtrsim 10 \fm$ for the deuteron bound state (filled red circles) and $L \gtrsim 8 \fm$ for the triton bound state (filled green circles). The first finite-volume excited state in the deuteron channel (empty red squares) or both the ground and the first excited state in the dineutron channel (empty blue circles and squares, respectively) correspond to the finite-volume scattering states. Their $|\Delta B_N|$ behavior, measured this time to the free-space two-body separation threshold $B_2^{\rm free}=0$, does not follow Eq.~(\ref{LuschBoundform}) but instead can be in the large $L$ asymptotic region expanded into the $\sim 1/L^{\beta}$ contributions \cite{ManZweGov83}.

\begin{figure}
  \includegraphics[width=8.6cm]{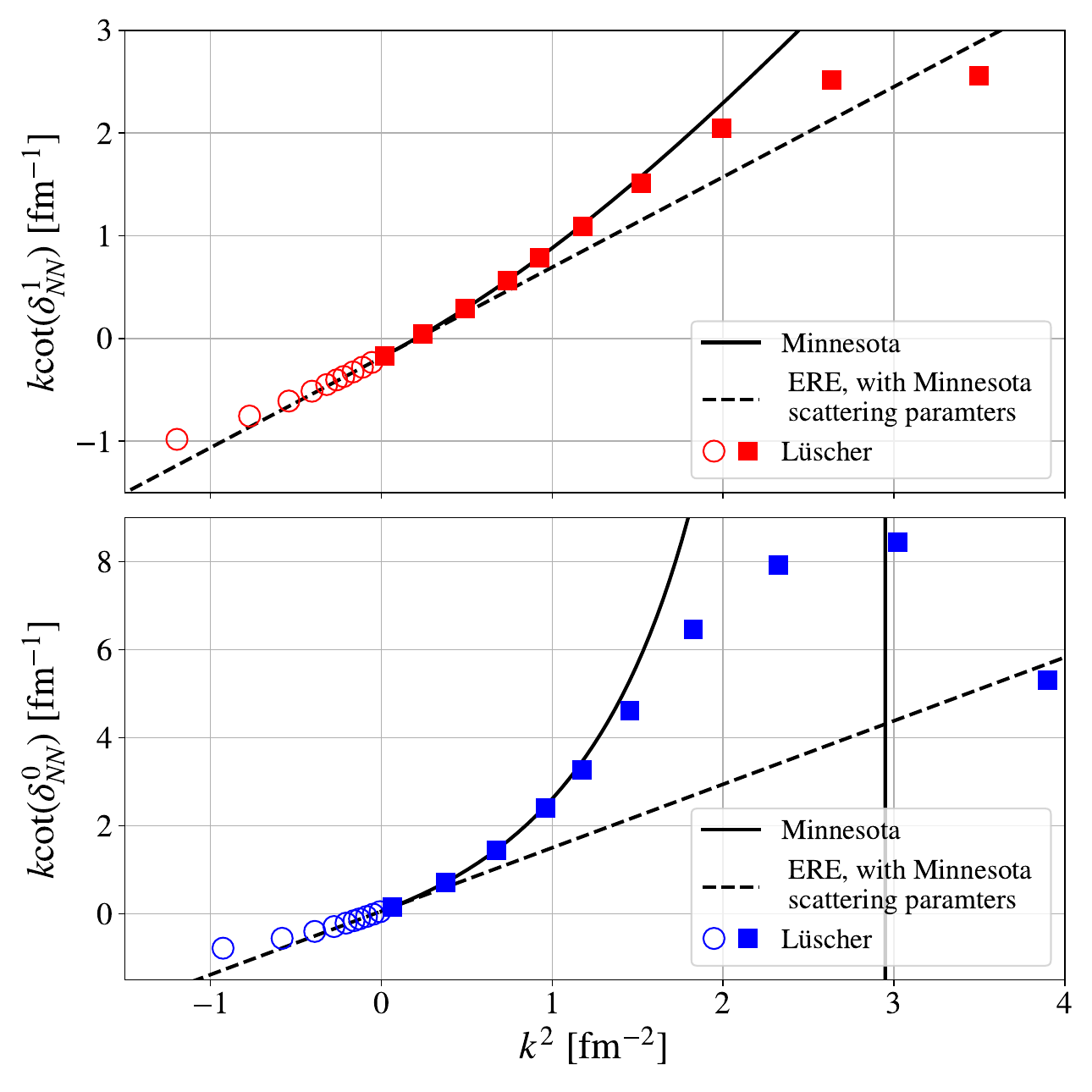}
  \caption{The $NN$ $s$-wave phase shifts, presented as $k\cot(\delta_{NN})$, for the deuteron (upper panel) and dineutron (lower panel) channels plotted as a function $k^2$. We compare the phase shifts calculated directly from the Minnesota potential (black solid line) to the phase shifts extracted from the Minnesota finite-volume energies using the L\"{u}scher scattering formula, Eq.~(\ref{LuschScattForm}). The phase shifts were calculated from both the finite-volume ground-state (empty circles) and the first-excited-state (filled squares) energies using different box sizes $L=\{3,3.5,4,4.5,5,5.5,6,7,9,20\}$~fm. For reference, the black dashed lines show ERE, Eq.~(\ref{eq: ERE }), with the first two terms fitted to the "free-space" Minnesota phase shifts (black solid lines). 
  }
\label{fig:phase_shifts}
\end{figure}

The $s$-wave phase shifts are extracted from the calculated two-body $A_1^+$ Minnesota finite-volume energies at various box sizes $L \in \langle3;20 \rangle~{\rm fm}$ using the L\"{u}scher scattering formula, Eq.~(\ref{LuschScattForm}). Fig.~\ref{fig:phase_shifts} shows the corresponding values in terms of $k {\rm cot}(\delta_{NN})$ as a function of $k^2$, which were calculated using the finite-volume ground state and the first excited state energies in the deuteron and dineutron channels. While the finite-volume energies of the first excited state are positive, the attractive nature of the $NN$ Minnesota potential introduces negative $A_1^+$ finite-volume ground-state energies in both two-body channels \cite{ManZweGov83,BeanBedParSav}. By approaching the free-space limit ($L \rightarrow \infty$) the first excited state energies converge to $B_2^{\rm free}=0$ from above, the spin-singlet ground-state energy converges to $B_2^{\rm free}=0$ from below, and the spin-triplet ground-state energy converges to the free-space deuteron energy. Eq.~(\ref{LuschScattForm}), used at different box sizes, yields $k {\rm cot}(\delta_{NN})$ at different $k^2>0$ (first excited states) or $k^2<0$ (ground states). The latter can be understood as the analytical continuation of the phase shifts to the momenta below the two-body threshold. Once the size of the box starts to approach the range of Minnesota interaction, the results obtained from the L\"{u}scher scattering formula begin to deviate from the Minnesota values calculated directly in free space (black solid lines). This is apparent for phase shifts at $k^2 \gtrsim 1.5 ~{\rm fm^{-2}}$ which are extracted from the finite-volume first excited state energies at box sizes $L\lesssim 5$~fm.           

\nopieft is limited to momenta up to the one pion-exchange threshold, $k \sim m_{\pi}$. Eq.~(\ref{LuschScattForm}) relates $A_1^+$ two-body finite-volume states with energies $E(L)$ to the free-space $s$-wave phase shifts with the momenta $|k|=\sqrt{m |E(L)|}$. As a result, we can identify which two-body finite-volume energy levels introduce high-momentum physics outside the scope of the theory and are not admissible to fit the \nopieft LECs. 
For the Minnesota potential, the two-body finite volume energies of the first excited state begin to surpass the breakdown scale of \nopieft for box sizes $L\lesssim 7~{\rm fm}$ (deuteron channel) and $L\lesssim 8~{\rm fm}$ (dineutron channel). The same is observed at smaller box sizes $L \lesssim 4~{\rm fm}$ for the two-body finite volume ground-state energiesin both the deuteron and dineutron channels. The three-body LECs are fitted to the $A_1^+$ ground-state finite-volume energies in the triton channel. Assuming that the box size $L$ is sufficiently large and the finite-volume correction to the free-space deuteron binding energy can be neglected, $E_d(L)\simeq E_d$, the three-body finite-volume energies $E_t(L)$ below the free-space three-nucleon separation threshold can be related to the $s$-wave neutron-deuteron ($nd$) phase shifts with momenta $|k_{nd}|\simeq \sqrt{4/3 m |E_t(L) - E_d|}$ \cite{RomShaBla19}. This suggests that by fitting the three-body LECs to $E_t(L)$ at $L \lesssim 8$~fm, we might start to introduce high momenta that exceed the breakdown scale of our theory.      

To emulate usual LQCD input scenarios and to demonstrate an option how to bypass possible high-momenta of two-body finite-volume excited states, we employ two distinct approaches to fix the \nopieft parameters up to NLO:
\begin{enumerate}
\item The fitting procedure is performed by considering the finite-volume energies generated only for one size of the box $L$. At LO, we fit the $C_{0}^{(0)}(\Lambda),\; C_{1}^{(0)}(\Lambda),$ and $D^{(0)}_{0}(\Lambda)$ LECs to reproduce exactly the Minnesota ground-state finite-volume energies in the dineutron, deuteron, and triton channels, respectively. The NLO two-body LECs $C_{0}^{(1)}(\Lambda)$, $C_{1}^{(1)}(\Lambda)$, $C_{2}^{(1)}(\Lambda)$, and $C_{3}^{(1)}(\Lambda)$ are perturbatively adjusted using the LO wave functions to further reproduce the Minnesota first excited state finite-volume energies in the respective two-body spin-isospin channels, while keeping the ground-state energies at the same LO values. The remaining $D_0^{(1)}(\Lambda)$ NLO LEC is perturbatively adjusted to keep the correct reproduction of the Minnesota finite-volume ground-state energy in the triton channel. 

\item We employ information only from the ground-state finite-volume levels and no excited state energies are considered. At LO, we fix again the $C_{0}^{(0)}(\Lambda),\; C_{1}^{(0)}(\Lambda),$ and $D_0^{(0)}(\Lambda)$ LECs by using the calculated Minnesota finite-volume energies in the dineutron, deuteron, and triton channels, respectively. However, this time, we fit each LO LEC by applying a $\chi^2$ fit to best match two Minnesota finite-volume ground-state energies which have been calculated for two different box sizes $\avgL\pm  1\fm$ in the corresponding channel. At NLO, the two-body LECs $C_{0}^{(1)}(\Lambda)$, $C_{1}^{(1)}(\Lambda)$, $C_{2}^{(1)}(\Lambda)$, and $C_{3}^{(1)}(\Lambda)$ are perturbatively adjusted to reproduce the Minnesota finite-volume ground-state energies in both $\avgL\pm  1\fm$ boxes exactly. Finally the $D_0^{(1)}(\Lambda)$ LEC is perturbatively fixed to keep the LO $\chi^2$-fitted three-body finite-volume ground-state energy at the same LO value. 
\end{enumerate}

If not written otherwise, we fit \nopieft for each value of momentum cutoff $\Lambda$ using different box sizes $L \in \{3, 5, 7, 9, 11, 13, 15\}\fm$ (first approach) and $\avgL \in \{4, 6, 8, 10, 12, 14\}\fm$ (second approach). To assess the cutoff dependence we consider several $\Lambda \in \{1.25, 2, 4, 6, 8, 10\} \fm^{-1}$ values.  

\section{Results}

Using the two different fitting scenarios described above, we present in this section the \nopieft predictions for both the deuteron, triton free-space binding energies and the low-energy $s$-wave $NN$ spin-singlet, $NN$ spin-triplet, and $nd$ spin-quartet scattering parameters. Where it is possible, we benchmark the \nopieft results by employing the L\"{u}scher formalism where we limit ourselves to the same finite-volume energies as has been considered during the \nopieft fit. The resulting free-space quantities are then shown as a function of the box size and are compared to the free-space results calculated directly from the Minnesota potential.  

\subsection{Deuteron channel \label{subs:deuteron}}

We start with the free-space deuteron bound state energy. The corresponding value calculated directly from the Minnesota potential is 
$$E_d \simeq - 2.202\,\mathrm{MeV}.$$

\begin{figure}
  \includegraphics[width=8.6cm]{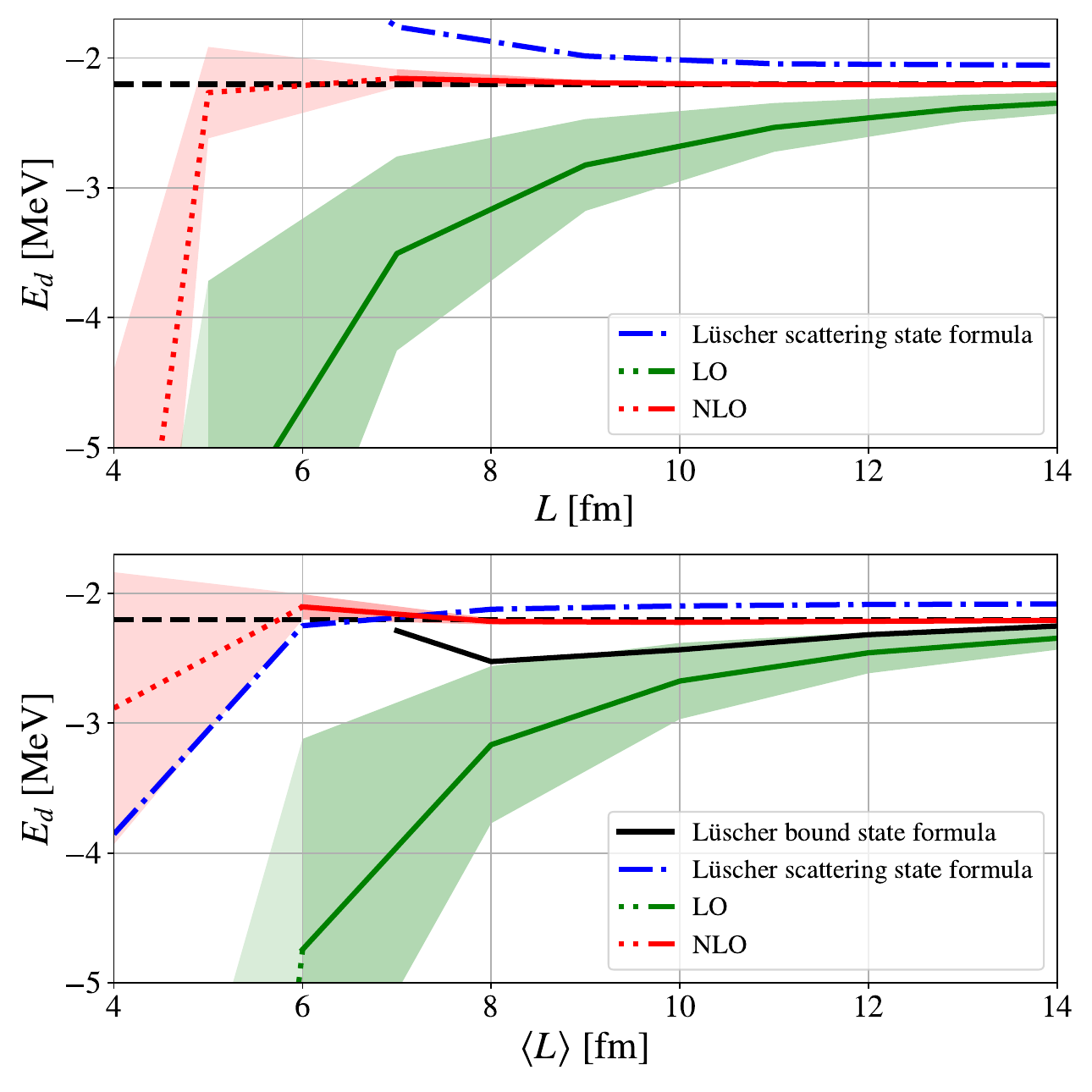}
  \caption{The free-space deuteron binding energy $E_d$ extracted from finite-volume energies is plotted against the varying box size $L$ employed in the extraction. 
  We compare different methods:
  In the upper panel, we use the ground-state and the first-excited-state energies from the same box size $L$. We show the results from Eq.~(\ref{LuschScattForm}) (blue dash-dotted line) and \nopieft at LO (green) and NLO (red) with LECs constrained through the first fitting scenario. 
  In the lower panel, we use the finite-volume ground state energies from two adjacent box sizes $\avgL \pm 1$~fm. Shown are the results obtained by fitting Eq.~(\ref{LuschBoundform}) (black solid line), Eq.~(\ref{LuschScattForm}) (blue dash-dotted line), and \nopieft at LO (green) and NLO (red) with LECs constrained through the second fitting scenario. The shaded bands in our \nopieft results indicate the uncertainty induced by the residual cutoff variation within the cutoff range $1.25 \leq \Lambda \leq 10~{\rm fm}^{-1}$. The lighter shaded bands with dotted lines reveal \nopieft results where at least one LEC is fitted to a finite-volume energy which might introduce momenta above the one-pion-exchange threshold. The result calculated directly from the Minnesota potential is shown as a black dashed line.}
  \label{fig:d_preds}
\end{figure}

In the first \nopieft fitting scenario, we employ the finite-volume energy of the ground state (LO) and the first excited state (NLO) evaluated using the Minnesota potential at one size of the box $L$. The \nopieft free-space deuteron bound state energy $E_d$ is then calculated by using the fitted LO and NLO LECs and solving the Schr\"{o}dinger equation outside the box. The upper panel of Fig.~\ref{fig:d_preds} presents the corresponding LO (green) and NLO (red) results as a function of $L$ employed during the fit. There is a significant improvement in our NLO results compared to the LO \nopieft prediction. For $L \gtrsim 7\fm$ the NLO \nopieft yields free-space $E_d$ energies fairly close to the Minnesota value. In this box size region, the residual cutoff variation (red shaded band) suggests that the NLO theoretical error is $\sim 3\%$. This surpasses the expected \nopieft NLO truncation error, which can be estimated as $(r^1_{NN}/a^1_{NN})^2\simeq 10\%$. A possible explanation is another small parameter at play, which is related to the relevant short-range scale divided by the box size. To put it differently, the NLO LECs are calibrated using deuteron binding energy in boxes significantly larger than the short-range scales not yet considered in the EFT order, this physics is essentially already accounted for in these LECs, and no additional truncation error is introduced.

In the second fitting scenario, we consider two finite-volume ground state energies from adjacent box sizes $\avgL \pm 1$~fm. The resulting free-space LO and NLO \nopieft deuteron binding energies, $E_d$, are shown for different $\avgL$ in the lower panel of Fig.~\ref{fig:d_preds}. Similar to the first case, once the NLO \nopieft is fitted to $\avgL \gtrsim 6\fm$ finite-volume input data, it successfully reproduces the deuteron free-space Minnesota binding energy.

The lighter-shaded bands with dotted lines depict the free-space deuteron energies, which were obtained using the \nopieft with LECs fitted to the finite-volume input data that introduce momenta above the one-pion-exchange threshold (see discussion in Subsec.~\ref{fitting}). The $E_d$ results start to deteriorate quite rapidly moving deeper into the small-$L$ region. Furthermore, our results confirm a slight advantage of the second fitting scenario which is renormalized only to the finite-volume ground-state energies and is not affected by the high momenta first excited state, which emerges already at $L \lesssim 7$~fm.

We move now to the L\"{u}scher approach. Due to the limitations imposed by the first fitting scenario, which involves finite-volume energies of the ground and first excited state within the same box, it is not possible to extract the free-space $E_d$ via Eq.~(\ref{LuschBoundform}). On the other hand, within the second fitting scenario, we have at our disposal deuteron finite-volume ground state energies for two different $L$. As a result, we can use Eq. (\ref{LuschBoundform}) and fit $\kappa_2$ and $\mathcal{A}_2$ to extract $E_d$. Interestingly, the Eq. (\ref{LuschBoundform}), when utilized with energies from two adjacent box sizes ($\avgL\pm 1\fm$), provides a solution only for $\avgL \gtrsim 7\fm$ and larger (see solid black line in the upper panel of Fig.~\ref{fig:d_preds}). To get an accuracy of about $5\%$ one has to use $\avgL \gtrsim 12\fm$. 

Another option is to employ L\"{u}scher's scattering state formula, Eq.~(\ref{LuschScattForm}). From the finite-volume ground state and the first excited state energies at the given $L$ (first fitting scenario) or two adjacent ground state energies at the average box size $\avgL$ (second fitting scenario), we extract the $s$-wave spin-triplet phase shifts at two different momenta. The $E_d$ is then obtained by fitting the ERE, Eqs.~(\ref{eq: ERE },\ref{b_e_scatt_params}), to these two phase-shift points.
These $E_d$ values are shown using the blue dash-dotted line in both the upper and lower panel of Fig.~\ref{fig:d_preds}. In the first case, the L\"{u}scher's scattering formula yields $E_d$ energies with $\sim 7 \%$ accuracy at $L \gtrsim 11\fm$. In the second case, the accuracy of $5\%$ is achieved when average box sizes $\avgL \gtrsim 8 \fm$ are used. At larger boxes, the accuracy does not improve due to the truncation of the ERE, Eq.~(\ref{eq: ERE }). Using a larger amount of finite volume excited states energies and considering higher ERE terms would likely further improve the $E_d$ result.

\begin{figure}
\includegraphics[width=8.6cm]{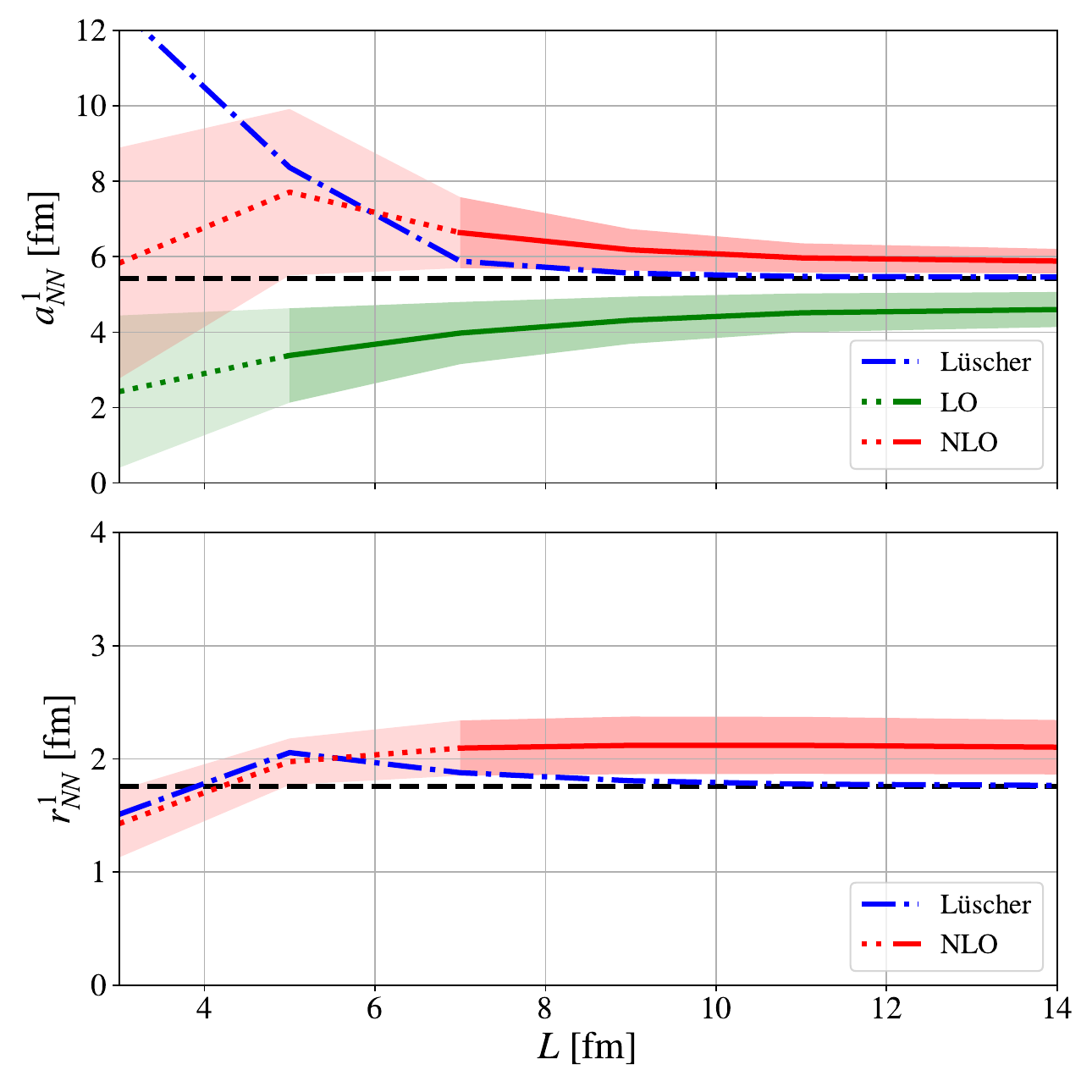}
    \caption{Same as the upper panel of Fig.~\ref{fig:d_preds} but for the $NN$ spin-triplet scattering length (upper panel) and effective range (lower panel).}
    \label{fig:3S1}
\end{figure}

The \nopieft results for $s$-wave spin-triplet scattering parameters are shown here only for the first fitting scenario. The outcomes of the second option are similar and the corresponding results are deferred to Appendix~\ref{appB}. The $s$-wave scattering parameters calculated directly from the Minnesota potential are 
$$a_{NN}^1 \simeq 5.427\fm,\;\; r_{NN}^1 \simeq 1.758\fm\,.$$
Fig.~\ref{fig:3S1} shows the predicted $NN$ spin-triplet scattering length and effective range at LO (green) and NLO (red) \mbox{\nopieft}, as a function of the box size $L$ employed in the fit. The figure does not show the calculated LO effective range, since it trivially behaves as $\Lambda^{-1}$ converging to zero in the contact limit. The blue dot-dashed lines represent $s$-wave scattering parameters which have been extracted by applying L\"{u}scher's scattering formula, Eq.~(\ref{LuschScattForm}). Again, both the \mbox{\nopieft}results up to NLO and the L\"{u}scher predictions were obtained using the finite volume ground state and first excited state energies from the same box size $L$.

The \mbox{\nopieft}and L\"{u}scher approach yield accurate predictions for the $a^1_{NN}$ and $r^1_{NN}$ values when utilizing box sizes $L \gtrsim 7\fm$, although the L\"{u}scher results converges more rapidly than the \mbox{\nopieft}results, with the latter being accurate within the uncertainty band. At smaller box sizes the LO or NLO LECs are fitted to finite-volume energies which eventually introduce momenta above the breakdown scale of the theory (lighter shaded bands with the dotted lines). As might be expected this leads to deterioration of our \nopieft results.

\subsection{Dineutron channel}

The $s$-wave scattering parameters in the $S,I=0,1$ dineutron channel are extracted from the finite-volume energies in a similar way as in the deuteron channel.
The corresponding Minnesota scattering length and effective range values are
$$a_{NN}^0 \simeq -16.80\fm,\;\; r_{NN}^0 \simeq 2.885\fm\,.$$

Both the first and the second \nopieft fitting scenarios, outlined in Sec.~\ref{fitting}, yield comparable dineutron scattering predictions. Here, we discuss results related only to the first case; the results obtained through the second scenario are again included in Appendix~\ref{appB}. Fig. \ref{fig:1S0} shows the calculated LO (green) and NLO (red) inverse $NN$ spin-singlet scattering lengths and effective ranges as a function of the box size $L$ used throughout the fit. The inverse of the scattering length is selected to avoid divergences when approaching the unitary point. As seen in the figure, the results of \nopieft demonstrate again a significant improvement from LO to NLO. Starting at the box size $L \simeq 7\fm$ and larger, the NLO \mbox{\nopieft}provides predictions close to the Minnesota values with accuracy of $\sim 1 \%$. In comparison, the size of the NLO theoretical error is roughly estimated as $(r^0_{NN}/a^0_{NN})^2\simeq 3\%$. The difference between the errors can be attributed to the presence of an additional small parameter (see Subsec.~\ref{subs:deuteron}).

The L\"{u}scher formalism, Eq.~(\ref{LuschScattForm}), accurately predicts the scattering length for $L \gtrsim 11\fm$, but it does not achieve the same level of accuracy for the effective range (blue dot-dashed lines). This can be explained by referring to Fig.~\ref{fig:phase_shifts}, where it is evident that an accurate description of the phase shifts in this spin channel with the ERE requires consideration of the shape parameter term. Therefore, it seems that the reduced accuracy in the L\"{u}scher prediction for the effective range is mainly due to the truncation of the ERE at second order in $k$.

\begin{figure}
\includegraphics[width=8.6cm]{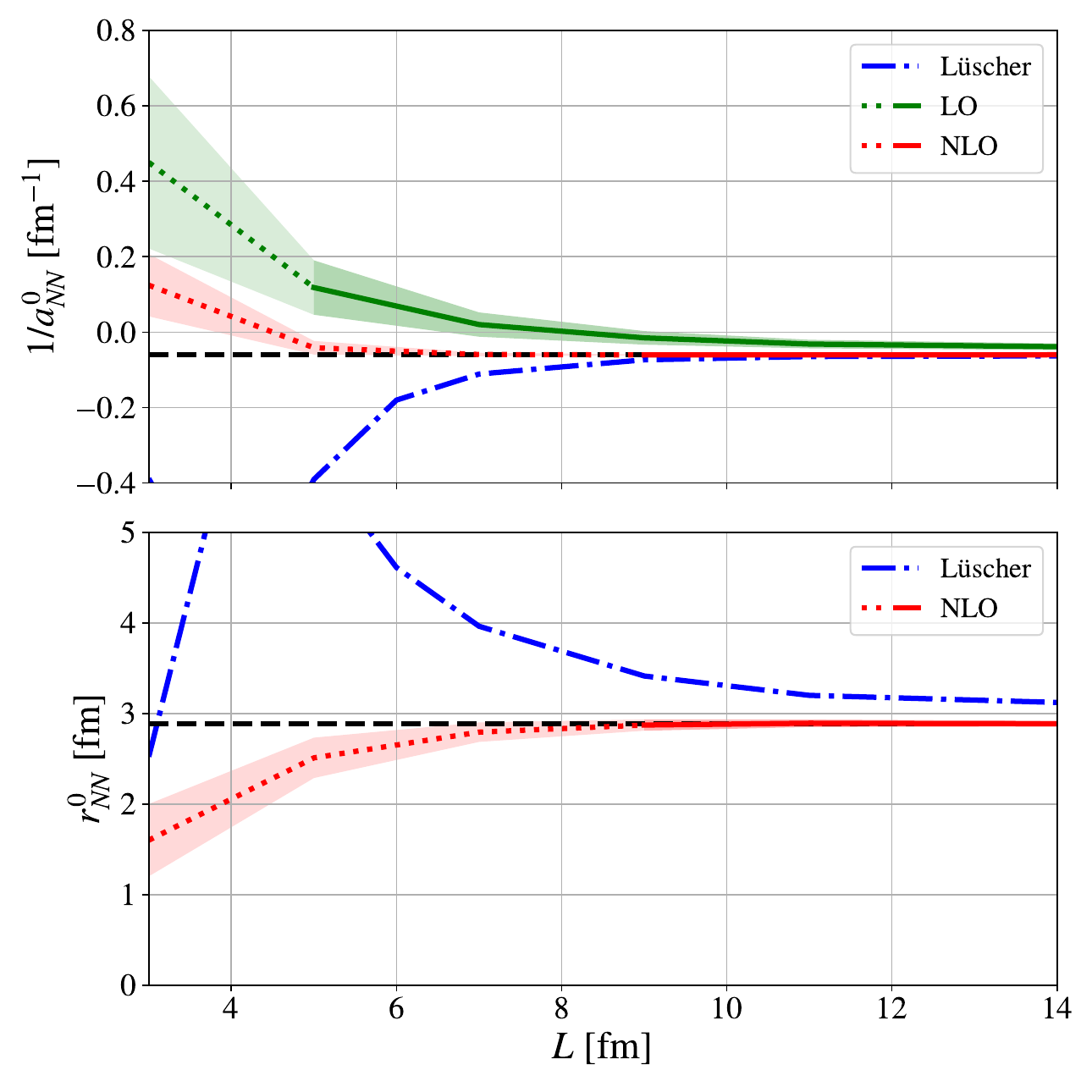}
  \caption{Same as the upper panel of Fig.~\ref{fig:d_preds} but for the $NN$ inverse spin-singlet scattering length (upper panel) and effective range (lower panel).}
  \label{fig:1S0}
\end{figure}

\subsection{Triton channel}

The three-body $S,I=1/2,1/2$ channel supports the triton bound state. Its free-space energy, as calculated directly from the Minnesota potential, is
$$E_t \simeq -8.385\,\mathrm{MeV}.$$

\begin{figure}
  \includegraphics[width=8.6cm]{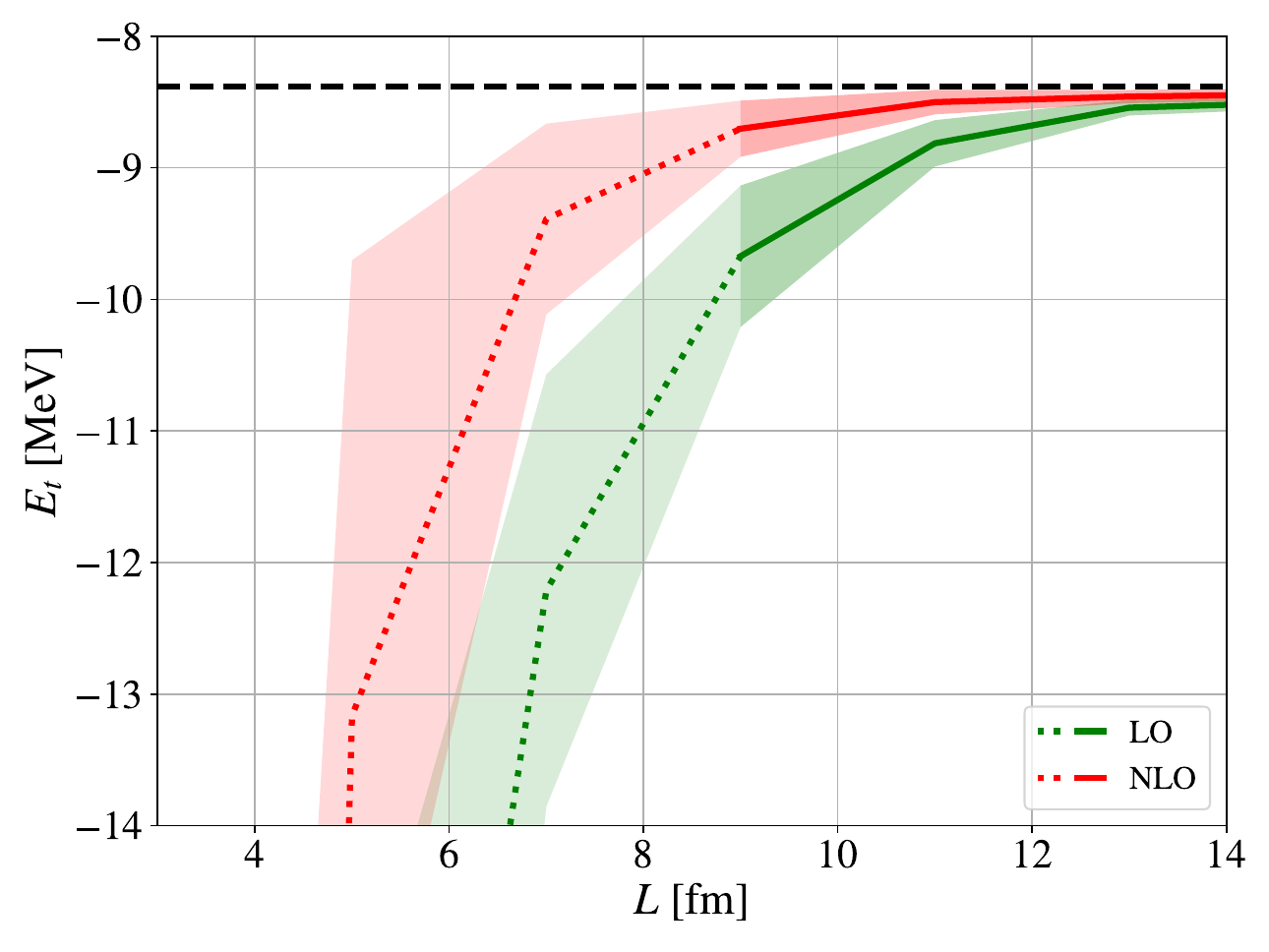}
  \caption{Same as the upper panel of Fig.~\ref{fig:d_preds} but for the free-space triton bound state energy $E_t$.}
\label{fig:t_preds}
\end{figure}

There is no new three-body term at NLO \nopieft and only a single three-body energy is needed to calibrate the theory. In line with our two-body studies, we show here free-space $E_t$ energies which have been calculated by using the \nopieft with LECs constrained through the first fitting scenario. The $E_t$ results obtained from the fit at two adjacent box sizes are given in Appendix~\ref{appB}. The resulting free-space energy, $E_t$, is shown in Fig.~\ref{fig:t_preds} as a function of the box size $L$ employed in the LEC fit. The NLO \nopieft results approach the Minnesota triton ground state energy with increasing $L$ and they reach an accuracy of $\sim 4 \%$ when box size $L = 9 \fm$ is utilized.

The triton calculation employs the full set of two- and three-body LO and NLO LECs all of which fitted to the finite-volume data which correspond to the relatively low free-space momenta. Since the finite-volume triton ground-state energy begins to exceed the breakdown scale of the \nopieft for $L \lesssim 8$~fm (roughly estimated in Sect.~\ref{fitting}), the three-body LECs fit is the most limiting factor in our finite-volume study. In this small $L$-range the calculated $E_t$ energies, both at LO and NLO, quickly diverge with decreasing $L$ from the free-space Minnesota value (see dotted lines with lighter shaded bands in Fig.~\ref{fig:t_preds}). As the box size gets smaller, also the two-body LECs start to gradually introduce momenta above the breakdown scale of the theory. This probably leads to a more pronounced and faster deterioration of our three-body results than in the previous two-body studies.

The lowest breakup threshold of the triton is into deuteron and neutron. Consequently, if one treats the deuteron as a point-like particle, Eq.~(\ref{NbodBound}) holds. 
Thus, in principle, having two triton finite-volume ground state energies at two different box sizes one can constrain the parameters $\kappa_3$ and $\mathcal{A}_3$ in the equation and extract the free-space $E_t$. This is the case in the second fitting scenario. The $E_t$ results obtained through this fit in two adjacent box sizes can be found enclosed in Appendix~\ref{appB}. There, the corresponding \nopieft triton results as well as free-space $E_t$ energies obtained using the L\"{u}scher-like formula, Eq.~(\ref{NbodBound}) are depicted in Fig.~\ref{fig:triton_b_e_gses}. 
It can be seen that the NLO \mbox{\nopieft}and the L\"{u}scher-like results converge very similarly, and both provide accurate results when average box sizes $\bra L\ket \gtrsim 10\fm$ are used.

\subsection{Three-nucleon $S=3/2, I=1/2$ channel}

The $S, I=3/2,1/2$ three-body channel does not support a bound state, and so we focus solely on the $s$-wave $nd$ scattering. The scattering parameters calculated directly from the Minnesota potential are
$$a_{nd}^{3/2}\simeq 6.327\fm,\;\;r_{nd}^{3/2} \simeq 1.968\fm.$$

This channel serves as a noteworthy demonstration of a significant advantage of the EFT. While the L\"{u}scher approach can relate the finite-volume results to the corresponding free-space quantity and can not predict properties of different systems based on results from another one, the EFT offers a more versatile solution. By calibrating the EFT using empirical data, it possesses the predictive power to address different characteristics of a variety of bound or unbound nuclear systems.
Using fitted \nopieft up to NLO, which has been employed up to now, no further finite-volume results are required, and the properties of the $s$-wave $nd$ spin-quartet scattering are a pure prediction.

To this end, we solve the $nd$ system with the harmonic potential added to the \nopieft potential, resulting in a spectrum of bound states. Scattering information can then be extracted from the fitting condition relating the phase shifts of the $s$-wave $\delta^{3/2}_{nd}(k)$ to the corresponding bound-state energy spectrum in the trap \cite{Busch1}
\be k \cot \left [ \delta^{3/2}_{nd}(k) \right] = - \sqrt{4~\mu_{nd}~\omega}~\frac{ \Gamma \left( 3/4 - \epsilon_\omega / 2\omega \right) }{ \Gamma \left( 1/4 - \epsilon_\omega / 2\omega \right)}.
\label{eq:busch}
\ee
Here $k$ stands for the relative $nd$ momentum and $\mu_{nd} \simeq 2m/3$ is the respective reduced mass. $\epsilon_\omega = E_\omega(nd)-E_\omega(d)$ is the energy of the trapped $nd$ state with respect to the $d$ energy threshold. 
Taking the limit of vanishing trapping potential, one can minimize the trap effect and obtain pure free-space results. This method has been used before, for example in Refs.~\cite{RSB12, SB23, BSBB23}, and has the advantage of obtaining scattering properties with a method intended for bound state calculations.

\begin{figure}
 \includegraphics[width=8.6cm]{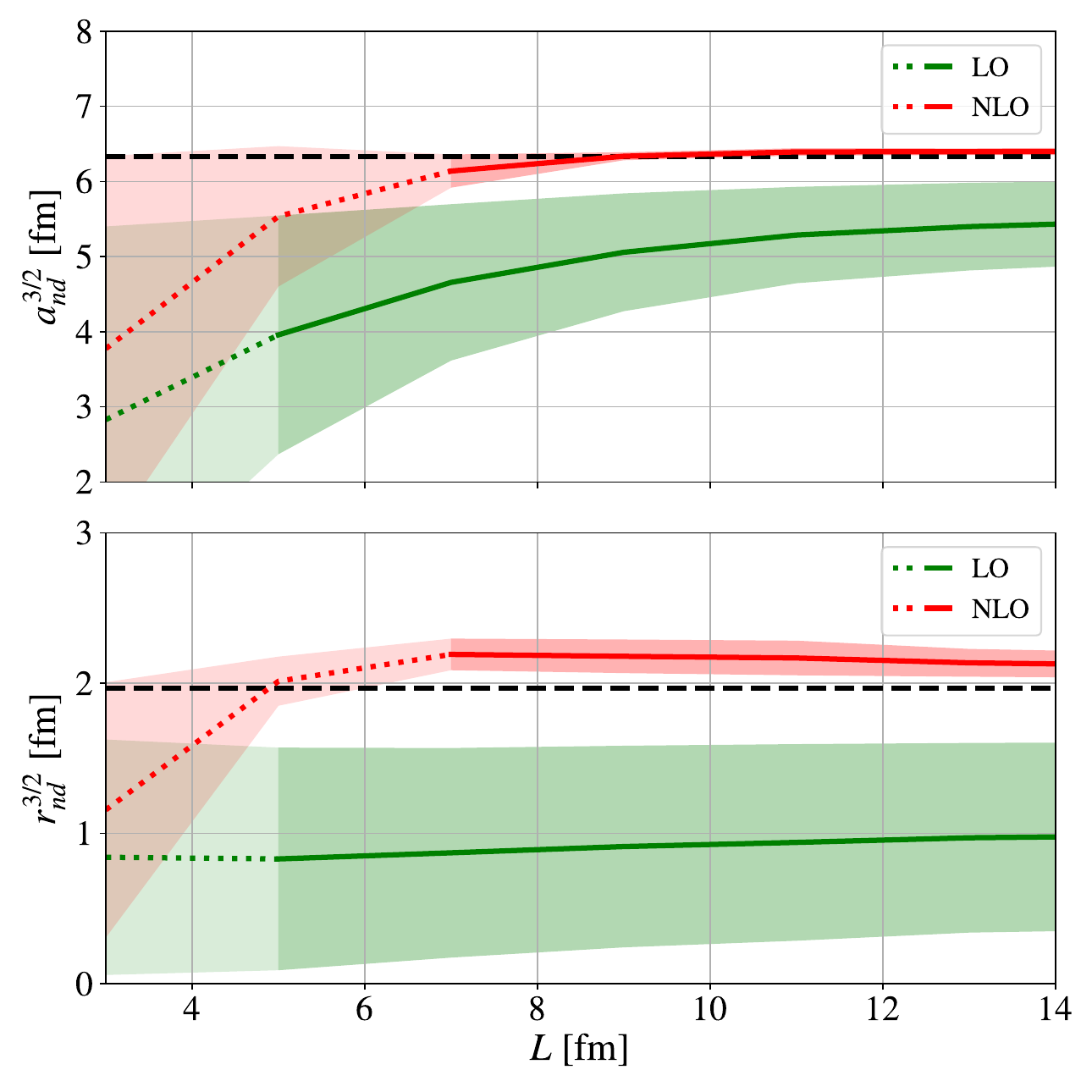}
 \caption{Same as the upper panel of Fig.~\ref{fig:d_preds} but for the $nd$ spin-quartet scattering length (upper panel) and effective range (lower panel).}
 \label{fig:nd_scatt_params_preds}
\end{figure}

The \nopieft results for the $nd$ spin-quartet scattering parameters are shown in Fig.~\ref{fig:nd_scatt_params_preds}. It is apparent that the inclusion of NLO terms significantly decreases the residual cutoff dependence and shifts predicted $a_{nd}^{3/2}$ and $r_{nd}^{3/2}$ parameters closer to the exact Minnesota values. Only spin-triplet LO and NLO terms contribute in the $s$-wave $nd$ spin-quartet channel. Consequently, the results depicted with the dotted lines reveal the same $L$-region, as in the Figs.~\ref{fig:d_preds}~and~\ref{fig:3S1}, where the LO or NLO LECs are fitted to the finite-volume energies corresponding to the free-space momenta above the one-pion-exchange threshold.

\subsection{Robustness of the Methods}

After obtaining the results described in the previous subsections, we evaluate the robustness of the L\"{u}scher and NLO \nopieft approaches by examining how the uncertainties in the artificial finite-volume LQCD input data affect the free-space results. In the first step, we assume a relative uncertainty of $5\%$ for the input finite-volume energies. In the second step, we propagate these uncertainties into our free-space results: For the L\"{u}scher approach, we use analytical derivatives, while for the \nopieft approach, we perform a Monte Carlo simulation. For simplicity, we restrict this study only to the two-body sector with \nopieft LECs fitted through the first scenario. We use one cutoff value $\Lambda = 4\fm^{-1}$ and we consider two different box sizes $L=5\fm$ and $7\fm$.

In Tab.~\ref{tab:d_nn}, we summarize calculated free-space deuteron bound state energy, scattering lengths, and effective ranges with the propagated uncertainties given in the parenthesis. Using LECs fitted in the $L = 5\fm$ finite-volume, the \nopieft $E_d$ result has a relative uncertainty of $30\%$. Considering the same $L = 5\fm$ finite-volume energies, the L\"{u}scher result has a relative uncertainty of about $85\%$. On the other hand, for $L = 7\fm$, both the L\"{u}scher result and the \nopieft predictions for the spin-triplet scattering length and effective range show a relative uncertainty of about $9\%$. There is an exceptional case in the dineutron channel for $L = 5 \fm$, where the \nopieft prediction has a larger relative uncertainty compared to the L\"{u}scher. Nevertheless, both uncertainties remain substantial and exceed $50\%$. Based on the table, it is visible that the L\"{u}scher results have either larger or comparable relative uncertainty to the corresponding NLO \nopieft predictions.

\begin{table}
\begin{center}
\caption{The two-body free-space results in the deuteron and dineutron channels. We compare the results calculated directly from the Minnesota potential to those extracted from boxes of size $L=5\fm$ and $7\fm$ using the L\"{u}scher formulas, Eqs.~\ref{LuschBoundform}~and~\ref{LuschScattForm}, and utilizing NLO \nopieft. The errors in the parenthesis correspond to propagated finite-volume input data uncertainties (for a detailed explanation see the text)}
\label{tab:d_nn}
\begin{tabular}{c c c c c c} 
 \hline\hline
 & &\multicolumn{2}{c}{ $L = 5\fm$ } & \multicolumn{2}{c}{$L = 7\fm$} \\
                & Minnesota & L\"{u}scher & NLO  & L\"{u}scher & NLO \\ 
 \hline
 $E_d$ [MeV]       & -2.2 & -0.8(7) & -2.5(8) & -1.8(4)& -2.2(4) \\ 
 $a_{NN}^{1}$ [fm] & 5.4  & 8.4(3.1)&  6.1(8) & 5.9(5) &  6.3(6) \\
 $r_{NN}^{1}$ [fm] & 1.7  & 2.1(3)  &  1.8(1) & 1.9(2) &  2.0(2) \\ 
 $1/a_{NN}^{0} [\fm^{-1}]$ & -0.058 & -0.4(3)  & -0.02(2) & -0.11(5) & -0.05(2)\\
 $r_{NN}^{0}   [\fm]$      &  2.885 & 5.8(3.2) &  2.3(2)  &  4(1)    &  2.7(3) \\
 \hline\hline
\end{tabular}
\end{center}
\end{table}

The propagated uncertainty in the L\"{u}scher predictions can be attributed to the behavior of the $S(\eta)$ function in Eq.~(\ref{S(eta)}) that governs the prediction. Within the $\eta$ values relevant to this study, $S(\eta)$ changes rapidly, which means that even small deviations from the exact value lead to different predictions. On the other hand, the \nopieft seems to be less sensitive to the accuracy of the input data. It can be inferred that the \nopieft might be in general more robust than the L\"{u}scher approaches, especially when it comes to the finite-volume input data uncertainties.

\section{Conclusions}

The primary objective of this work is to explore the NLO \nopieft with perturbative effective range corrections fitted to finite-volume input data - a crucial consideration for forthcoming LQCD calculations pertinent to nuclear physics. To achieve this, we employ both the free space and finite volume versions of the SVM, with the latter incorporating periodic boundary conditions. These few-body techniques provide us with a highly accurate framework to address the finite-volume effects at varying box sizes as well as in the free-space limit. In lieu of nuclear LQCD results at physical pion mass, and to facilitate a quantitative comparison with established data, we emulate the finite-volume spectra by solving the Minnesota $NN$ potential.

From these finite-volume data, we extract the free-space binding energies and $s$-wave scattering parameters for two- and three-nucleon systems. We then juxtapose the outcomes obtained through the L\"{u}scher methods with those derived from \nopieft at both LO and NLO. The free-space results, which are obtained by applying directly the Minnesota potential, serve as the benchmark.

Our findings reveal a substantial improvement of \nopieft results when transitioning from LO to NLO. Moreover, the NLO \nopieft generally provides predictions close to the free-space Minnesota values when using the box sizes $L \gtrsim 7\fm$. In most two-body scenarios, the NLO \nopieft effectively captures the finite-volume effects and incorporates the necessary corrections. This trend is also evident in the $nd$ $S= 3/2$ channel, which at NLO involves only a two-body scale. By introducing relative uncertainties into the finite-volume input data, we demonstrated the robustness of \nopieft results. On the other hand, the L\"{u}scher free-space results remain rather sensitive to the propagated uncertainties.   

For $L \lesssim 5\fm$, the L\"{u}scher formalism yields the $NN$ spin-singlet and spin-triplet $s$-wave phase shifts, which deviate from the corresponding free-space Minnesota values. This suggests that the finite-volume corrections in this small $L$-region start to be dominated by the strong interplay between the size of the box and the range of the Minnesota potential. The same might potentially affect any EFT with LECs fitted at such small box sizes. In our study, the finite-volume energies in this $L$-region start to correspond, through the L\"{u}scher formalism, to $NN$ phase shifts with momenta above the breakdown scale of \nopieft. Here, our results quickly deteriorate with decreasing $L$ and they should be considered as a mere extrapolation outside the scope of the theory. This sets a hard limit on the validity of our \nopieft predictions.

\nopieft proves to be a robust tool, not only due to its capacity to yield precise free-space results but also because it enables the extraction of meaningful insights with limited finite-volume input data. This results in a reduced number of unknown parameters that must be determined. In contrast, the L\"{u}scher formalism demands finite volume energies from the same system to compute free space observables, while EFT offers the potential to transcend this requirement. 

In future \nopieft works, it would be interesting to explore higher orders of the theory, which introduce $p$-wave, $d$-wave, and non-central interaction terms. This would require the study of $T_1^-$, $E^+$, and $T_2^+$ finite-volume spectra, where the leading contribution in the $T_1^-$ cubic symmetry is induced by the $p$-wave, while for the $E^+$ and $T_2^+$ cubic symmetries it is the $d$-wave. Furthermore, it is highly topical to use EFTs to explore the LQCD finite-volume results with non-zero strangeness. Here, recent advances in LQCD baryon-baryon and meson-baryon calculations \cite{HALQCD1,HALQCD2,Bulava23} seem to converge rather fast to the physical pion mass. Using these LQCD results would provide an intriguing insight into a strangeness sector, where the amount of low-energy experimental data is often limited.             

\section*{ACKNOWLEDGMENT}
We would like to thank Nir Barnea and Sebastian K\"{o}nig for useful discussions and communications. The work of M. Sch\"{a}fer was supported by the Czech Science Foundation GA\v{C}R grant 22-14497S.

\appendix
\renewcommand\thefigure{\thesection.\arabic{figure}}    
\section{Finite-volume data \label{appA}}
The finite-volume input data are generated using the Minnesota $NN$ potential \cite{Minn} with mixing parameter $u=1$. In Tab.~\ref{tab:data}, we list the calculated energies of the two lowest laying states of the $A_1^+$ deuteron and dineutron spectra for different $L$. The finite-volume ground-state energies in the triton channel are given as well. In our calculations, we use the nucleon mass parameter $(\hbar c)^2/m \simeq 41.471 ~{\rm MeV.fm^2}$.

\begin{table}
\begin{center}
\caption{Finite-volume input data which are employed to constrain the \nopieft LO and NLO LECs. Listed are the ground-state ($E_0$) and the first excited state ($E_1$) $A_1^+$ energies in the deuteron and dineutron channels as well as the $A_1^+$ ground-state energy value in the triton channel, which are calculated at different box sizes $L$. The last row of the table corresponds to the free-space results.}
\label{tab:data}
\begin{tabular}{c r r r r r} 
 \hline\hline
 \multirow{2}{*}{$L$ [fm]}& \multicolumn{2}{c}{Deuteron} & \multicolumn{2}{c}{Dineutron}& \multicolumn{1}{c}{Triton} \\
                & \multicolumn{1}{c}{$E_0$} & \multicolumn{1}{c}{$E_1$} & \multicolumn{1}{c}{$E_0$}  & \multicolumn{1}{c}{$E_1$} & \multicolumn{1}{c}{$E_0$} \\ 
 \hline
  3& -49.6002& 145.0738 & -38.4197 & 161.5753 & -131.7462\\ 
  5& -13.2650&  48.8501 &  -8.5695 &  60.2897 &  -32.8364\\
  7&  -6.8728&  20.3600 &  -3.6048 &  27.9261 &  -16.5174\\ 
  9&  -4.5895&  10.1385 &  -1.9741 &  15.5872 &  -11.3680\\
 11&  -3.5205&   5.6454 &  -1.2355 &   9.8703 &   -9.4258\\
 13&  -2.9523&   3.3794 &  -0.8396 &   6.8068 &   -8.7281\\
 15&  -2.6326&   2.1290 &  -0.6036 &   4.9852 &   -8.4973\\ \hline
$\infty $& -2.2019 & 0 & 0  &  0   & -8.3850  \\
 \hline\hline
\end{tabular}
\end{center}
\end{table}

\section{Additional results \label{appB}}
We introduced two different options of how to fit LO and NLO \nopieft LECs to finite-volume data (see Sec.~\ref{fitting}).
The first scenario uses the finite-volume ground state and the first excited state energy values at the same single box size $L$, while the second scenario considers only the finite-volume ground state energies from two adjacent box sizes $\avgL \pm 1 \fm$. For free-space deuteron bound state energy $E_d$ we show the corresponding \nopieft results of both fitting methods in Sec.~\ref{subs:deuteron}. For completeness, we enclose in this appendix additional results obtained with the second fitting scheme using two adjacent box sizes.

The results of $NN$ $s$-wave scattering parameters are shown for spin-triplet, Fig. \ref{fig:3S1_2BOX}, and spin-singlet, Fig. \ref{fig:1S0_2BOX}. L\"{u}scher's scattering formula, Eq.~(\ref{LuschScattForm}), can also be used here, and its results are shown as well. The free-space triton bound state energy is given in Fig. \ref{fig:triton_b_e_gses}. In the same figure, we show the results of the generalization of L\"{u}scher's bound state formula, Eq.~\ref{NbodBound}, as well. Finally, the $s$-wave $nd$ spin-quartet scattering parameters are presented in Fig. \ref{fig:nd_scatt_params_gses}. 

\setcounter{figure}{0}
\begin{figure}
\includegraphics[width=8.6cm]{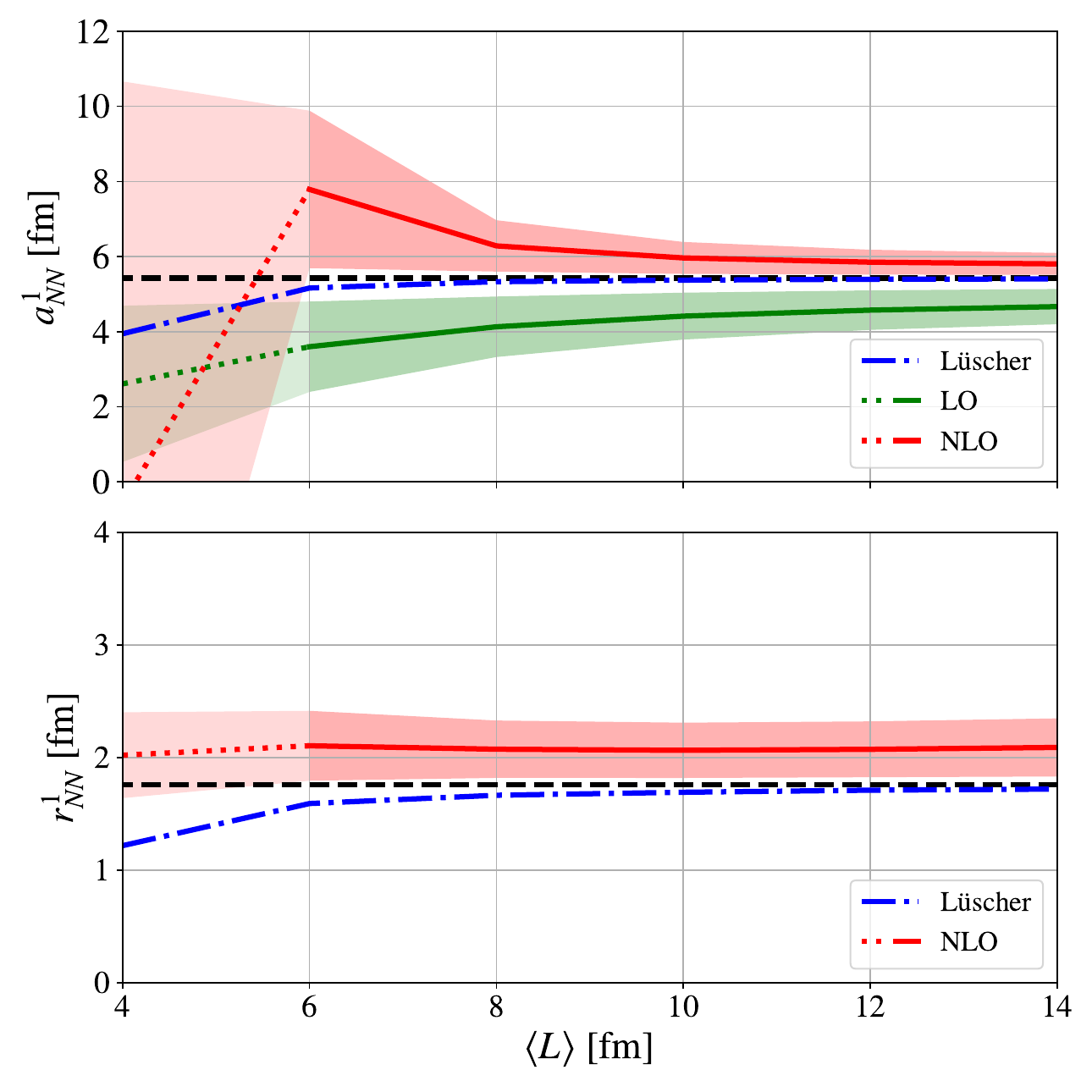}
 \caption{Same as the lower panel of Fig.~\ref{fig:d_preds} but for the $NN$ spin triplet scattering length (upper panel) and effective range (lower panel).}
    \label{fig:3S1_2BOX}
\end{figure}

\begin{figure}
\includegraphics[width=8.6cm]{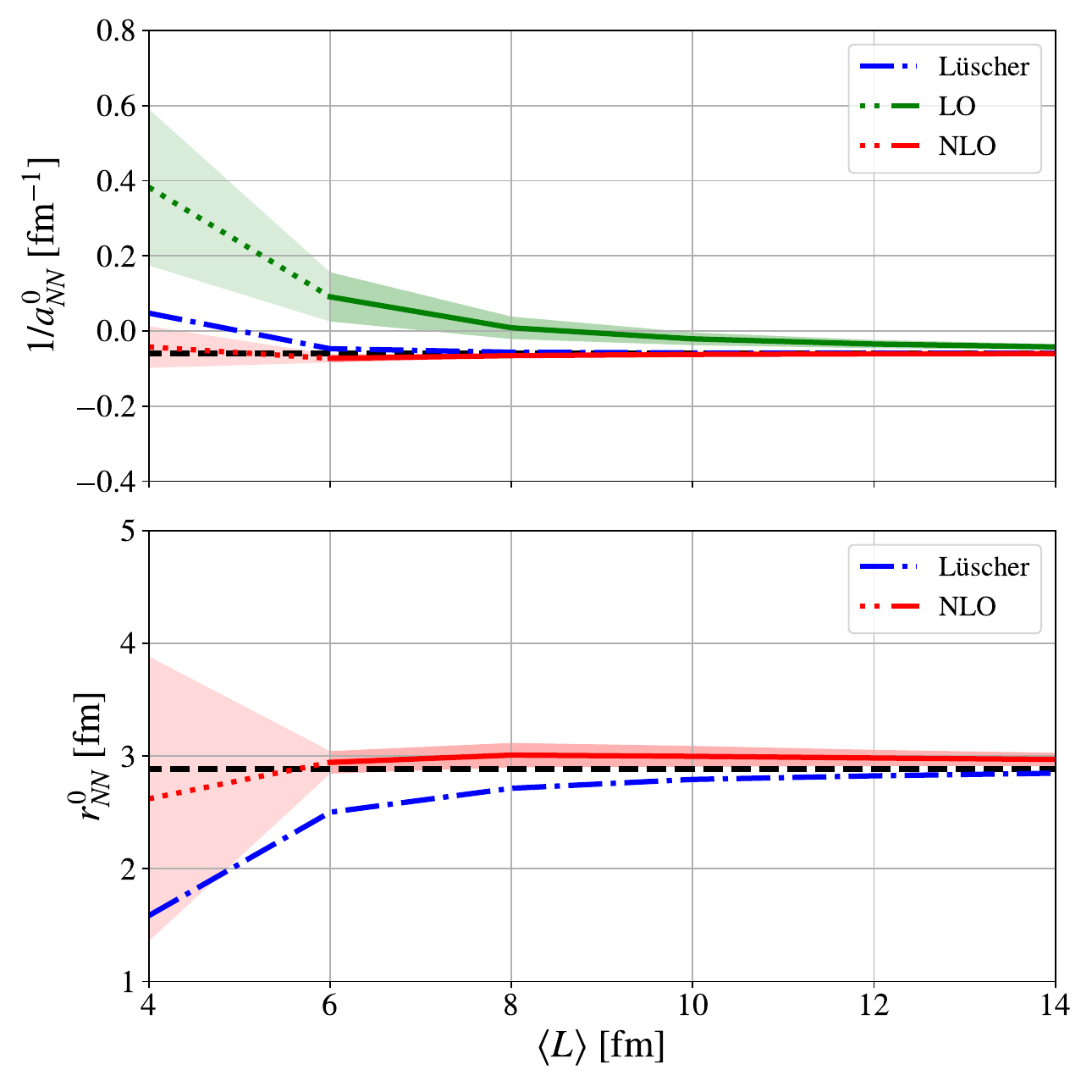}
    \caption{Same as the lower panel of Fig.~\ref{fig:d_preds} but for the $NN$ inverse spin-singlet scattering length (upper panel) and effective range (lower panel).}
    \label{fig:1S0_2BOX}
\end{figure}

\begin{figure}
    \centering
    \includegraphics[width = 8.6 cm]{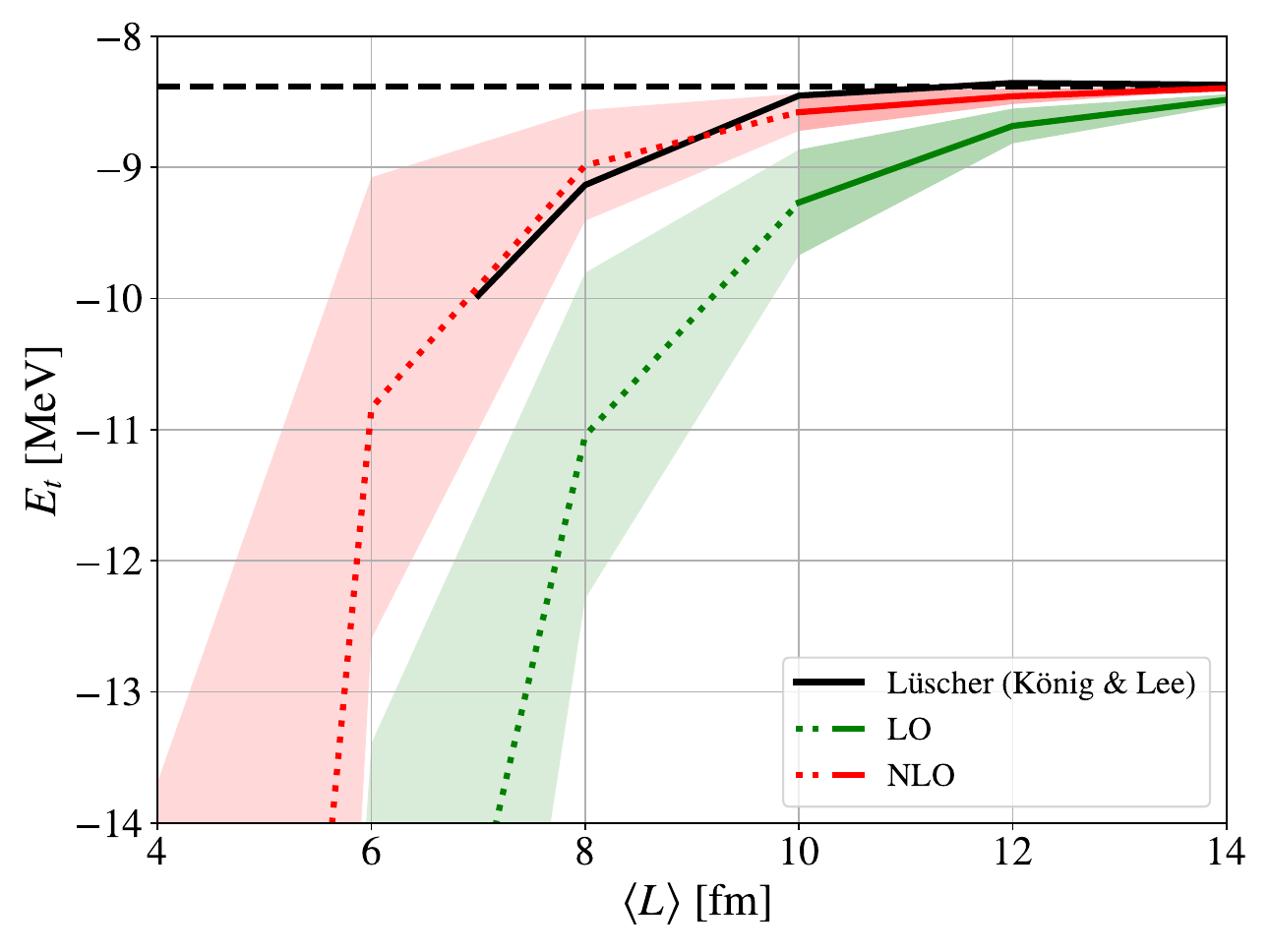}
    \caption{Same as the lower panel of Fig.~\ref{fig:d_preds} but for the free-space triton ground state energy $E_t$. Here, the black solid line show $E_t$ results obtained through the generalized L\"{u}scher bound state formula, Eq.~(\ref{NbodBound}).}
    \label{fig:triton_b_e_gses}
\end{figure}

\begin{figure}
    \centering
    \includegraphics[width = 8.6 cm]{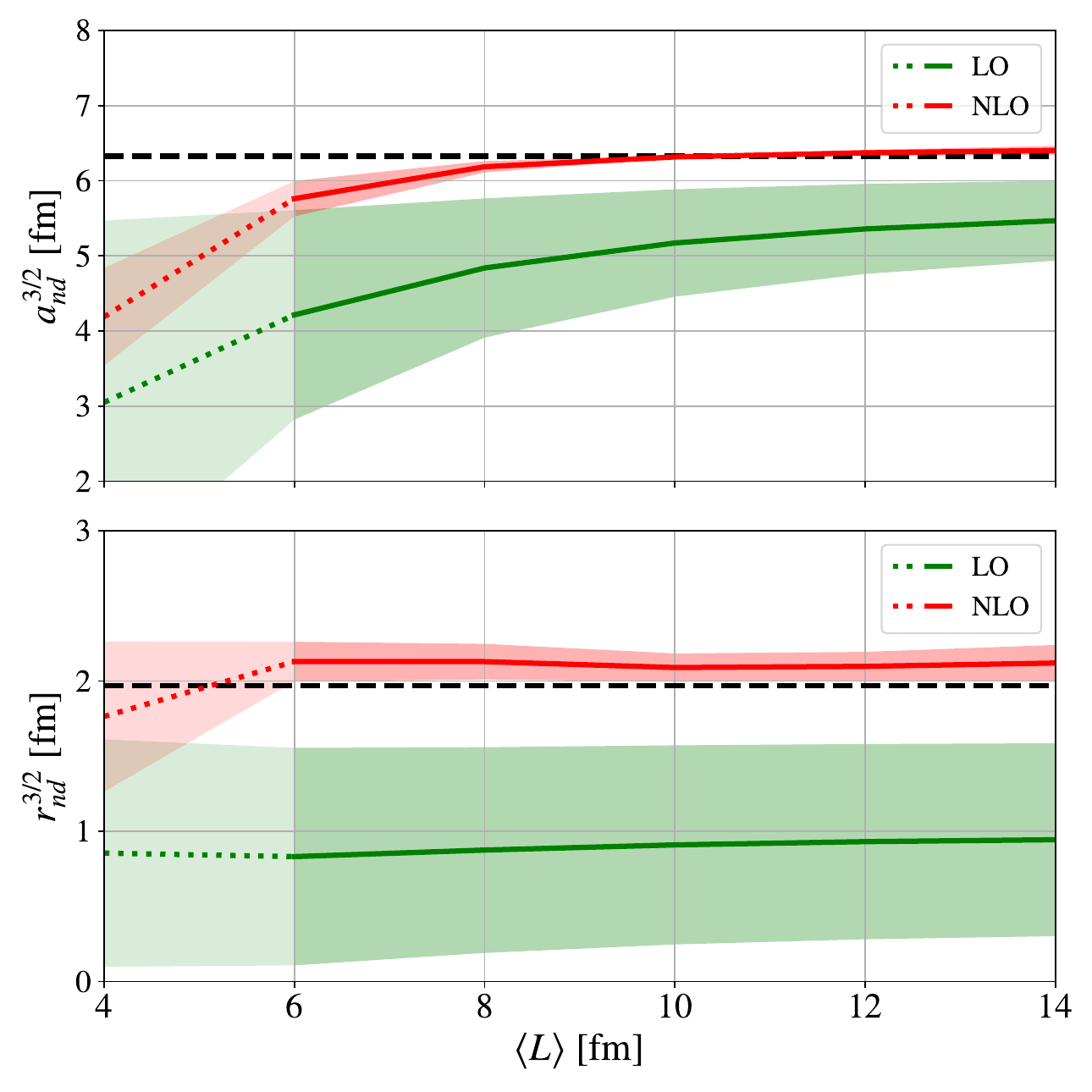}
    \caption{Same as the lower panel of Fig.~\ref{fig:d_preds} but for the $nd$ spin-quartet scattering length (upper panel) and effective range (lower panel).}
    \label{fig:nd_scatt_params_gses}
\end{figure}


\end{document}